\documentclass[12pt,]{article}

\usepackage{newtxtext,newtxmath}

\usepackage{graphicx}
\usepackage{parskip}
\usepackage{comment}
\usepackage{lineno}

\usepackage[letterpaper,margin=1in]{geometry}

\renewenvironment{abstract}
	{\quotation}
	{\endquotation}

\date{}

\makeatletter
\renewcommand{\fnum@figure}{\textbf{Figure \thefigure}}
\renewcommand{\fnum@table}{\textbf{Table \thetable}}
\makeatother

\usepackage{scicite}

\usepackage{url}

\def\scititle{Ultrafast electron crystallography reveals the atomic pathway of a light-driven correlated insulator-to-metal transition}

\title{\bfseries \boldmath \scititle}

\author{
A.~Niedermayr$^{1\ast^\dagger}$,
H.~Xu$^{2,3\ast^\dagger}$,
C.~S.~Ong$^{4^\dagger}$,
P.~Thunström$^{4}$,
\and
M.~Yannai$^{5}$,
J.~Wu$^{1}$,
G.~Cao$^{1}$,
L.~Kornblum$^{5}$,
I.~Kaminer$^{5,6}$,
\and
O.~Grånäs$^{4}$,
X.~Zou$^{2}$,
J.~Weissenrieder$^{1\ast}$
\and
\small$^{1}$Light and Matter Physics, Applied Physics,\\
\small School of Engineering Sciences, KTH Royal Institute of Technology, Stockholm SE-100 44, Sweden.
\and
\small$^{2}$Department of Chemistry, Stockholm University, Stockholm, Sweden.
\and
\small$^{3}$Research School of Chemistry, Australian National University, Canberra, Australia.
\and
\small$^{4}$Department of Physics and Astronomy, Uppsala University, Box 516, 75120 Uppsala, Sweden.
\and
\small$^{5}$Andrew and Erna Viterbi Department of Electrical and Computer Engineering,\\
\small Technion--Israel Institute of Technology, Haifa 32000, Israel.
\and
\small$^{6}$Department of Materials Science and Engineering,\\
\small Technion--Israel Institute of Technology, Haifa 32000, Israel
\and
\small$^\dagger$These authors contributed equally to this work.\\
\small$^\ast$Corresponding authors. Email: arthurn@kth.se; Hongyi.Xu1@anu.edu.au; jonas@kth.se.
}

\begin{document}
\iftrue
\maketitle

\begin{abstract} \bfseries \boldmath
Ultrafast phase transitions in correlated materials are often inferred from selected diffraction peak intensities or diffraction peak displacements, leaving the underlying three-dimensional atomic trajectories elusive. Resolving these trajectories is essential for identifying which atomic motions drive changes in electronic properties and how they couple to electronic degrees of freedom. We address this challenge in vanadium dioxide ($\mathrm{VO_2}$), a correlated oxide with a near-room-temperature transition between the insulating monoclinic (M1) phase and metallic rutile (R) phase. For this purpose, we introduce ultrafast three-dimensional electron diffraction, which enables refinement of the transient unit cell and internal V and O coordinates, revealing the V-V dimerization and zigzag motion during the phase transition. The refined atomic coordinates follow a linear trajectory in real space during the transition, in contrast to nonlinear or sequential transformation pathways inferred from more indirect observables in earlier work. Quantum many-body calculations treating each V–V pair as a correlated unit show that dimerization creates the level splitting responsible for the electronic gap, which is further enhanced by nonlocal electronic interactions between the paired V atoms. The gap collapses when dimerization is lost. This work turns time-resolved diffraction from order-parameter tracking into transient crystallography, directly connecting atomic trajectories to electronic mechanisms in correlated materials.
\end{abstract}

Ultrashort laser pulses can transiently reshape the electronic, magnetic, and
structural properties of materials, in some cases creating states that cannot be stabilized
under equilibrium conditions~\cite{stojchevska_ultrafast_2014,
vogelgesang_phase_2018, horstmann_coherent_2020}. Such capabilities have opened
routes to ideas such as light-induced superconductivity~\cite{fausti_light-induced_2011}, all-optical
switching of magnetic order on picosecond timescales~\cite{stanciu_all-optical_2007}, and metastable structural phases that underpin commercial phase-change memory~\cite{zalden_femtosecond_2019}.
Nonetheless, determining the atomic structures and transition pathways of these short-lived states has
remained beyond the reach of full crystallographic refinement from ultrafast diffraction data.
The limitation is intrinsic: obtaining complete, high-precision diffraction datasets on ultrafast timescales is challenging because transient scattering signals are weak and measurements must balance temporal resolution, reciprocal-space coverage, acquisition time, and sample damage. As a result, time-resolved electron and x-ray diffraction
experiments often track one or a few selected selected Bragg reflection intensities or structural order parameters ~\cite{baum_4d_2007,li_direct_2022,johnson_all-optical_2024}. This
contrasts with static electron and x-ray diffraction captured under equilibrium conditions, where extensive datasets of three-dimensional diffraction have enabled precise determination of
fractional atomic coordinates within the unit cell~\cite{ma_single-crystal_2018,
gruene_establishing_2021}.

In this work, we overcome this limitation by extending three-dimensional
electron diffraction (3DED) into the ultrafast regime, a technique we call
ultrafast 3DED~\cite{hennicke_3d_2026}. Static 3DED is an established method for atomic-scale structure
determination of sub-micron crystals across materials science and structural
biology, where single-crystal x-ray diffraction is often not
feasible~\cite{gemmi_3d_2019,clabbers_macromolecular_2021}. Its strength lies in
measuring Bragg intensities over a series of crystal tilt angles, thereby
sampling the crystal from multiple orientations relative to the electron beam.
We combine this tilt-series acquisition with an optical pump-electron probe
scheme (Fig.~\ref{fig1}a), accumulating diffraction signal stroboscopically over
many pump-probe cycles and merging Bragg intensities measured over the full tilt
series at each pump-probe delay. The resulting 3DED dataset is refined directly to obtain the transient atomic coordinates within
the unit cell during the photoexcited phase transition (Fig.~\ref{fig1}b). By acquiring diffraction data off the main-zone axes and integrating Bragg reflections over multiple tilt angles, ultrafast 3DED suppresses multiple scattering and improves intensity accuracy relative to conventional 2D diffraction measurements, allowing the average transient crystal structure to be determined at every pump--probe delay rather than inferred from changes in selected Bragg reflections.

We demonstrate 3DED on vanadium dioxide ($\mathrm{VO_2}$), a prototypical strongly correlated material, in which the atomic pathway of the photoinduced insulator-to-metal transition has remained unresolved despite
decades of study. This material undergoes a coupled
structural and electronic phase transition near 340~K, from a monoclinic (M1)
insulator with dimerized vanadium pairs to a tetragonal metallic rutile (R) phase
(Fig.~\ref{fig1}c). The M1 distortion relative to the rutile parent can be
decomposed into three coupled structural components: (i) a monoclinic shear,
reflected in the deviation of the monoclinic angle $\beta$ (defined between the
$a$ and $c$ lattice vectors) from the rutile-reference value of approximately
$116^\circ$ in the same monoclinic setting, corresponding to what is sometimes
described as unit-cell distortion; (ii) V--V dimerization along the chains of vanadium atoms running parallel to the rutile $c_R$ axis,
in which neighboring V atoms form alternating short intra-dimer and long
interdimer V--V distances; (iii) a lateral zigzag displacement, in which the V
atoms shift transversely to the chain in an alternating pattern, corresponding
to what is sometimes described as dimer tilting or chain twisting. These components
vanish in the ideal rutile structure and together define the monoclinic M1
distortion~\cite{baum_4d_2007, li_direct_2022,xu_transient_2023}. This
decomposition provides a structural framework for describing the photoinduced
M1-to-R pathway and for identifying the components of the M1 distortion driving
the electronic gap opening.

Recent experimental and theoretical studies suggest that the M1-to-R transition
may proceed through pathways more complex than a direct structural interpolation.
Here, a direct, linear pathway denotes a coordinated evolution in which the
atomic coordinates move along a straight line in three-dimensional space between the M1 and R structures. A nonlinear or asynchronous pathway instead denotes a
trajectory that deviates from this straight structural interpolation, either
because different components evolve on different timescales or because the
system passes through an intermediate crystallographic structure. Li~\textit{et
al.}~\cite{li_direct_2022} reported that V--V dimerization and lateral zigzag
initially evolve toward their rutile values along a linear, common-timescale
trajectory. At later delays, however, the motion no longer follows this
trajectory: the lateral zigzag continues to relax, whereas the V--V dimerization
changes comparatively little. Baum~\textit{et al.}~\cite{baum_4d_2007} inferred
a stepwise, temporally separated pathway: a femtosecond component assigned to
V--V bond dilation, followed by picosecond atomic displacements and slower shear
motion. Johnson~\textit{et al.}~\cite{johnson_all-optical_2024} found that
incoherent, spatially correlated lattice distortions, identified as interacting
polarons, create a transient state in which the energy barrier to the metallic
phase is reduced. Theoretical work by Grandi~\textit{et
al.}~\cite{grandi_unraveling_2020} proposed a two-step transition in which V--V
dimerization is lost before the lateral zigzag distortion, producing an
intermediate monoclinic metallic phase between the M1 insulator and the rutile
metal.


While these works significantly advanced our understanding of the mechanisms behind the ultrafast phase transition, they do not determine the complete transient crystallographic structure; instead, the transition pathway is inferred indirectly from projected coordinates, Bragg-intensity classes, diffuse-scattering correlations, or theoretical order-parameter landscapes. In contrast, ultrafast 3DED refines a structural model against Bragg intensities measured over a series of crystal tilt angles. Each pump--probe delay yields a three-dimensional dataset containing several hundred Bragg reflections, sufficient to refine the fractional coordinates of all symmetry-inequivalent atoms in the unit cell. This capability extends ultrafast electron diffraction beyond order-parameter tracking to time-resolved crystallographic refinement, enabling direct tracking of the atomic trajectories throughout the photoexcited phase transition.

In this work, we excite a structural phase transition in $\mathrm{VO_2}$ using
laser pulses with a photon energy of  1.2 eV and a duration of 300~fs. The resulting
changes in diffraction are probed using 1.5 ps electron pulses in a stroboscopic
approach at a repetition rate of 12 kHz. The sample is a focused ion beam (FIB)
lamella prepared from a single-crystalline $\mathrm{VO_2}$
micro-needle~\cite{fisher_moving_1975}, with dimensions of \(4 \times 5 \, \mu
\text{m}^2\) and an approximate thickness of 100 nm (see SM). The experimental data are processed using the REDp
software~\cite{wan_three-dimensional_2013}, which reconstructs the time-resolved 3D reciprocal
lattice. The intensities of the Bragg reflections in each 3DED dataset at a given time delay are integrated using XDS~\cite{kabsch_xds_2010}. These diffraction patterns can be projected along arbitrary zone axes
(Fig.~\ref{fig2}a). To monitor the phase transition, we track the total
intensity of the Bragg peaks (Fig.~\ref{fig2}b), splitting them into two
categories: M1-exclusive peaks, which are present in the monoclinic M1 phase but
absent in the rutile R phase, and shared peaks, common to both structures. The
M1-exclusive peaks are sensitive to component (ii) and (iii) of the M1 distortion. At each
delay, the signal is averaged over more than 500 such reflections. The M1-exclusive intensity follows a
double-exponential decay (Fig.~\ref{fig2}b), with both the bi-exponential dynamics and the associated timescales consistent with previous reports~\cite{baum_4d_2007,johnson_ultrafast_2022, li_direct_2022}: a fast decay reduces the intensity to
approximately 50\% within $\sim10$~ps, and a slower decay completes the
suppression on a characteristic timescale of 400~ps. The intensity loss at the
M1-exclusive peaks reflects the decrease of the dimerization amplitude. In the
M1 phase, the V atoms are organized into two interleaved sub-lattices displaced
from each other by the dimerization, with twice the rutile periodicity along the
chain. This doubling of the periodicity along the rutile $c_R$ direction produces the M1-exclusive peaks and suppresses the structure factor at the shared peaks. As the photoexcited transition
proceeds and the dimerization decreases, the two sub-lattices merge into a
single uniform lattice: the M1-exclusive peaks are suppressed, and the intensity of the shared
peaks grows. This anticorrelated trend is observed in our data, consistent with
the redistribution of Bragg intensity as the M1 superstructure is suppressed.

Beyond intensity tracking, ultrafast 3DED allows us to determine the unit-cell
parameters and the atomic fractional coordinates (the positions of atoms
expressed as fractions of the unit-cell dimensions) at each pump-probe delay. We use
this capability to track the lattice parameters of the unit cell, as well as the
individual atomic coordinates within it. Of particular interest for $\mathrm{VO_2}$ is the time-dependent evolution of the monoclinic angle $\beta$, the angle between the $a$ and $c$ lattice vectors of the unit cell (Fig.~\ref{fig2}c), which directly tracks component (i), the monoclinic shear. In contrast to the V--V dimerization and zigzag displacement (Fig.~\ref{fig2}d,e), which exhibit a pronounced fast response immediately following photoexcitation, the evolution of $\beta$ is dominated by a slower exponential relaxation. The dimerization and zigzag components also display a slower evolution on a comparable timescale. This indicates that the slow relaxation of components (i)--(iii) toward their rutile values proceeds on a common timescale rather than asynchronously, even though components (ii) and (iii) additionally exhibit a distinct ultrafast response immediately following photoexcitation. Although our measurements may not accurately capture small absolute changes in $\beta$ because of the limited number of tilt angles, they determine its temporal evolution with high precision and reveal a clear trend. The high precision
arises because the experimental conditions, such as tilt angles and laser
illumination, remain consistent across all delays. Static measurements report
$\beta = 115.2^\circ$ for the monoclinic phase and approximately $116^\circ$ for
the rutile phase in the same monoclinic representation~\cite{kucharczyk_accurate_1979,xu_transient_2023}.

We employ two complementary approaches to determine the transient atomic coordinates during the phase transition. First, we use the
program SHELXL~\cite{sheldrick_crystal_2015,hubschle_shelxle_2011} to refine the atomic coordinates of
the vanadium (Fig.~\ref{fig3}a) and two oxygen atoms (see SM) using the full reflection set, with
all monoclinic structures successfully converging. This provides the absolute crystallographic coordinates within the refined transient unit cell. Second, using only the M1-exclusive reflections, we apply a least-squares refinement to determine the relative, time-dependent atomic coordinates. This provides an independent determination of the transient M1 structural evolution that excludes reflections that may contain contributions from the rutile phase. Both approaches yield excellent agreement of the V coordinates~(\ref{fig3}), providing a consistency check on the extracted transient structural evolution. The refined V fractional coordinates capture both components (ii) and (iii) of the M1 distortion: the longitudinal V--V pairing and the transverse zigzag displacement, respectively. As shown in Fig.~\ref{fig3}a, both structural components evolve from their M1 values toward the rutile configuration with a common overall relaxation behavior. The most abrupt changes occur within the first few picoseconds, during which the atoms traverse approximately half of the total structural displacement. This initial motion is followed by a much slower evolution extending beyond 1 ns, during which the remaining displacement is gradually completed. The vanadium and oxygen coordinates therefore follow nearly straight trajectories from their M1 to their R positions, defining a direct, linear pathway in coordinate space, as illustrated by the correlation plot in Fig.~\ref{fig3}b. Additional correlation plots of the vanadium fractional coordinates are shown in Fig.~S7 in the SM. The $x$-components of the fractional coordinates of the O1 and O2 atoms are shown in Fig.~\ref{fig3}c,d, respectively. All remaining components and correlation plots are shown in Fig.~S2. A residual V--V dimerization of approximately 15\% of its initial M1-to-R distortion nevertheless persists into the nanosecond regime, indicating that a small fraction of the initial V--V dimerization displacement remains despite the average crystallographic structure being indexed as rutile. Potential phase heterogeneity~\cite{qazilbash_mott_2007} is not expected to significantly affect the determination of the M1 atomic positions. Fractional coordinates obtained from the least-squares refinement to M1-exclusive reflections (see SM) are in close agreement with those obtained using the full reflection set in SHELXL, which also includes reflections potentially shared with the rutile phase. This agreement indicates that inclusion of the potentially phase-overlapping reflections does not significantly affect the refined transient M1 fractional coordinates.

Having tracked all three components of the M1 distortion as a function of pump--probe delay, we next identify which structural distortion is responsible for the collapse of the electronic band gap. The insulating M1 phase arises from the interplay of structural symmetry breaking and electronic correlations. In particular, the formation of V--V dimers along the rutile chains splits the chain-aligned $d_\parallel$ orbitals into bonding and antibonding states, while the accompanying distortion of the $\mathrm{VO_6}$ octahedra shifts the relative energy of the $e_g^\pi$ states (Fig.~\ref{fig:theory_bs}a). Together, these distortions create the structural framework for the insulating M1 phase, while electronic correlations determine the magnitude and stability of the gap. The relative importance of the individual structural distortions, however, has remained under debate.

The three structural components affect the electronic states of $\mathrm{VO_2}$ in different ways. The monoclinic shear changes the unit-cell geometry and modifies the V--O--V bond angles, thereby altering the dispersion of the electronic bands. However, this distortion alone does not directly split the V-derived states responsible for the insulating character of the M1 phase. In contrast, V--V dimerization along the rutile chains creates alternating short and long V--V bonds. This changes the overlap between neighboring V orbitals and splits the chain-aligned $d_\parallel$ states into bonding and antibonding states, providing the structural basis for the electronic gap formation. The lateral zigzag distortion primarily modifies V--O hybridization and shifts the relative energy of the $e_g^\pi$ states, thereby influencing the size of the gap, but does not generate the bonding--antibonding splitting of the $d_\parallel$ manifold. These considerations suggest that V--V dimerization is the structural distortion most directly linked to the electronic gap opening, consistent with the Goodenough--Eyert picture~\cite{goodenough_two_1971,eyert_metal-insulator_2002, zhang_hidden_2024}. Nevertheless, determining the quantitative contribution of each distortion requires explicit calculations of the correlated electronic structure, as the M1 gap emerges from the interplay of structural changes and electronic interactions.

To quantify how the experimentally observed structural changes affect the electronic properties, we perform cluster dynamical mean-field theory (DMFT) calculations on the experimentally refined transient structures. The correlated subspace is constructed from V-centered $t_{2g}$ orbitals, with each V--V dimer treated as a correlated impurity cluster. The calculations include local Coulomb interactions and Hund coupling, together with a nonlocal exchange contribution between the two V atoms of the dimer (see SM). The impurity problem is solved using continuous-time hybridization-expansion quantum Monte Carlo (CT-HYB)~\cite{werner_continuous-time_2006, werner_efficient_2006, seth_triqscthyb_2016} within the all-electron DFT+DMFT framework implemented in Relativistic Spin Polarized toolkit (RSPt)~\cite{Wills2010}. Calculations are performed at $T=300$~K, corresponding to the experimental conditions. For the equilibrium M1 structure, this approach yields an insulating gap of $0.67$~eV, in excellent agreement with the experimentally reported gap of approximately $0.6$--$0.7$~eV from optical spectroscopy, photoemission, and transport measurements~\cite{verleur_optical_1968,berglund_electronic_1969,shin_vanadium_1990,koethe_transfer_2006} (Fig.~\ref{fig:theory_bs}b). The cluster-DMFT solution is paramagnetic, and does not impose static long-range magnetic order, while capturing singlet correlations within the V--V dimers. This is consistent with the absence of long-range magnetic order in the M1 phase at room temperature~\cite{pouget_electron_1975,huffman_insulating_2017}.

We next use the experimentally refined M1 structure as a reference to disentangle the structural and many-body contributions to the $d_{\parallel}$ bonding--antibonding splitting. The calculated splitting is approximately 3~eV (Fig.~4b). V--V dimerization provides the one-body contribution by creating contrasting hopping amplitudes across the short and long V--V bonds, while many-body effects are described by the full frequency-dependent intersite self-energy of the dimer. In cluster DMFT, this self-energy arises from the on-site interactions $U$ and $J$ together with an explicit intersite exchange interaction $V_{\mathrm{ex}}$, which is included as a full two-body term in the cluster impurity interaction tensor (see SM). At a fixed frequency, the intersite hopping and intersite self-energy enter with opposite signs for the bonding and antibonding combinations, thereby modifying their separation, whereas site-local terms contribute equally to both. With $V_{\mathrm{ex}}=-0.49$~eV together with $U$ and $J$, the calculated gap is 0.67~eV and the occupied $d_{\parallel}$ bonding peak lies 0.6~eV below the chemical potential. Retaining $U$ and $J$ but setting $V_{\mathrm{ex}}=0$, with all other parameters unchanged, the system remains insulating, but the gap decreases to approximately 0.14~eV and the bonding peak shifts to only 0.2~eV below the chemical potential. We then completely remove the interaction tensor, setting $U=J=V_{\mathrm{ex}}=0$, while retaining the same refined M1 geometry. In this one-body limit, the $d_{\parallel}$ splitting decreases to approximately 1.5~eV and the separation between the occupied $d_{\parallel}$ bonding states and the low-lying $e_g^{\pi}$ states is lost (Fig.~4c). Together with the shear-only control below, these results identify V--V dimerization as the dominant structural origin of the bonding--antibonding splitting, while showing that many-body interactions substantially enhance this splitting and stabilize the full insulating M1 state.

At the Hartree--Fock level, the first-order off-diagonal self-energy generated by $V_{\mathrm{ex}}$ maps onto the static intersite correction of DFT+$V$, giving $V=-2V_{\mathrm{ex}}=0.98$~eV; the factor of two is the spin sum in the exchange contraction~\cite{Haas2024} (see SM). Cluster DMFT is not restricted to this static limit: $V_{\mathrm{ex}}$ is sampled as a full two-body interaction, including opposite-spin spin-exchange terms, and therefore also contributes to the frequency-dependent many-body self-energy. The short V--V bond length, in turn, controls the orbital overlap between neighboring V atoms and therefore strongly influences both the one-body hopping and the many-body intersite self-energy. The hopping across the short bond decreases as the bond lengthens. Microscopically, the screened exchange interaction represented by $V_{\mathrm{ex}}$ is likewise expected to weaken with increasing V--V distance. In a simple atomic-orbital picture, the overlap decreases approximately as $e^{-r/L}$, where $r$ is the short V--V bond length and $L$ is an effective orbital decay length. As the exchange matrix element contains an orbital-overlap factor at each electron coordinate, $\lvert V_{\mathrm{ex}}\rvert$ is expected to decrease approximately as $e^{-2r/L}$. We verified numerically that the magnitude of the intersite self-energy decreases as $\lvert V_{\mathrm{ex}}\rvert$ is reduced. The structure itself further modifies the intersite self-energy through changes in hopping, hybridization, and the intersite density matrix, which decreases as the short and long V--V bonds equalize. Thus, as V--V dimerization weakens along the measured photoinduced structural trajectory (Fig.~2d), the lengthening of the short V--V bond reduces the one-body hopping and weakens the many-body intersite contribution to the bonding--antibonding splitting.

We further examine whether the monoclinic shear component alone can induce an insulating state within the noninteracting one-body limit, $U=J=V_{\mathrm{ex}}=0$. We construct a hypothetical shear-only structure by placing the rutile fractional coordinates in the monoclinic unit cell, without V--V dimerization or the lateral zigzag displacement. The $d_{\parallel}$ bonding--antibonding splitting is strongly reduced and the calculated spectral function is metallic (Fig.~4d). Although the shear lifts the symmetry-enforced degeneracy of the $d_{\parallel}$ bands at the rutile Z point (see SM), the resulting splitting is too small to open a gap. This shows that the symmetry breaking associated with the monoclinic cell is not sufficient to produce the insulating M1 electronic structure. For comparison, we also calculate the rutile structure corresponding to the long-delay state, in which all three M1 distortion components vanish. The residual band splitting observed for the shear-only structure then disappears and the $d_{\parallel}$ states recover their rutile symmetry, yielding a metallic spectral function consistent with the experimentally observed metallic steady state at delays beyond 1~ns (Fig.~2d).


Beyond the interactions within the V--V dimer, V--O hybridization contributes to the insulating gap through both one-body band alignment and ligand-mediated many-body effects. The choice of correlated orbitals determines how these ligand-mediated interactions are incorporated into the many-body treatment. For predominantly V-centered orbitals, an on-site Hubbard interaction alone is insufficient to reproduce the M1 gap: single-site DMFT and spin-polarized DFT+$U$ remain metallic for $U$ values up to 4~eV when the explicit intersite exchange interaction is omitted. When the correlated orbitals contain more O-ligand character, the nominal on-site interaction acts on hybridized V--O orbitals and therefore incorporates contributions that are nonlocal in an atom-centered representation, including screened V--O Coulomb interactions and ligand-mediated contributions to the effective V--V interactions. In this ligand-rich subspace, the $d_{\parallel}$ bonding--antibonding splitting increases and a gap opens even without the explicit intersite exchange interaction. V--O hybridization therefore affects the gap in two complementary ways: ligand-mediated interactions enhance the $d_{\parallel}$ bonding--antibonding splitting, and one-body V--O hybridization controls the relative alignment of the $d_{\parallel}$ and $e_g^{\pi}$ states. The ligand-rich construction, however, does not reproduce a systematic collapse of the gap when dimerization is removed (see SM); the main calculations therefore use the predominantly V-centered subspace.

These results show that the M1 gap cannot be explained by the structural $d_{\parallel}$ bonding--antibonding splitting alone. V--V dimerization provides the essential one-body framework. The full frequency-dependent intersite self-energy, generated by the on-site interactions and the explicit intersite exchange interaction, further increases the $d_{\parallel}$ bonding--antibonding splitting. Ligand-mediated V--O interactions also contribute to this splitting, while one-body V--O hybridization controls the relative alignment of the $d_{\parallel}$ and $e_g^{\pi}$ states. Together, these electronic effects stabilize the full insulating state. A complete description of these coupled structural and electronic effects would require a unified treatment of dynamical correlations and nonlocal screening and exchange, such as $GW+$DMFT~\cite{Biermann2003,Ayral2016,Choi2016,Tomczak2017}. Recent quasiparticle $GW$ studies of VO$_2$ likewise show that the M1 gap is strongly sensitive to lattice distortions~\cite{weber_role_2020}.


In conclusion, ultrafast 3DED enables transient crystallography of VO$_2$ by resolving the time-dependent evolution of the unit cell and all symmetry-inequivalent V and O positions. Within the frozen-snapshot equilibrium mapping used here, combining these measurements with cluster DMFT calculations on the experimentally refined transient structures shows that the calculated gap collapse follows primarily from the loss of V--V dimerization along the measured structural trajectory. As the dimerization weakens, the reduced $d_{\parallel}$ overlap diminishes the one-body bonding--antibonding splitting and weakens the structure-dependent intersite self-energy contributions that stabilize the equilibrium M1 phase. Monoclinic shear alone is insufficient to open a gap. V--O hybridization affects both the $d_{\parallel}$ splitting through ligand-mediated interactions and the relative alignment of the $d_{\parallel}$ and $e_g^{\pi}$ states. Together, experiment and theory identify M1 VO$_2$ as a correlated dimer insulator and demonstrate how ultrafast 3DED can connect atomic-scale structural dynamics with electronic functionality in correlated materials. The reported trajectory describes the average structure; local distortions invisible to Bragg diffraction may persist.

\section*{Acknowledgments}
We thank Prof. Bertina Fisher for providing the $\mathrm{VO_2}$ microneedles used to create the FIB lamellae in this study. Language polishing was assisted by ChatGPT (OpenAI). All content was critically reviewed and approved by the authors.
\paragraph*{Funding:}
This research was funded by the Knut and Alice Wallenberg Foundation (2012.0321, 2018.0104, 2019.0124), the Swedish Research Council (VR, 2018.00815), and through the ARTEMI national infrastructure (VR 2021-00171, VR 2022-03596 and Strategic Research Council (SSF) RIF21-0026). A.N. acknowledges funding from the Swiss National Science Foundation (SNSF) through project P500PT\_214469. H.X. acknowledges funding from the Australian Research Council for the Future Fellowship (FT240100378). OG and CSO acknowledge funding from VR (2025-06120  and 2025-06380). OG, CSO and PT acknowledge funding from European Research Council (ERC), Synergy Grant 854843-FASTCORR. Computational resources were provided by the National Academic Infrastructure for Supercomputing in Sweden (NAISS), funded by the Swedish Research Council. Jianyu Wu gratefully acknowledges the Chinese Scholarship Foundation (CSC) for a doctoral fellowship. L.K. is grateful for the support of the Israel Science Foundation (ISF Grant No. 1397/24).
\paragraph*{Competing interests:}
There are no competing interests to declare.
\paragraph*{Data and materials availability:}
All data are available in the manuscript or the supplementary materials.

\clearpage

\begin{figure}%
\centering
\includegraphics[width=0.95\textwidth]{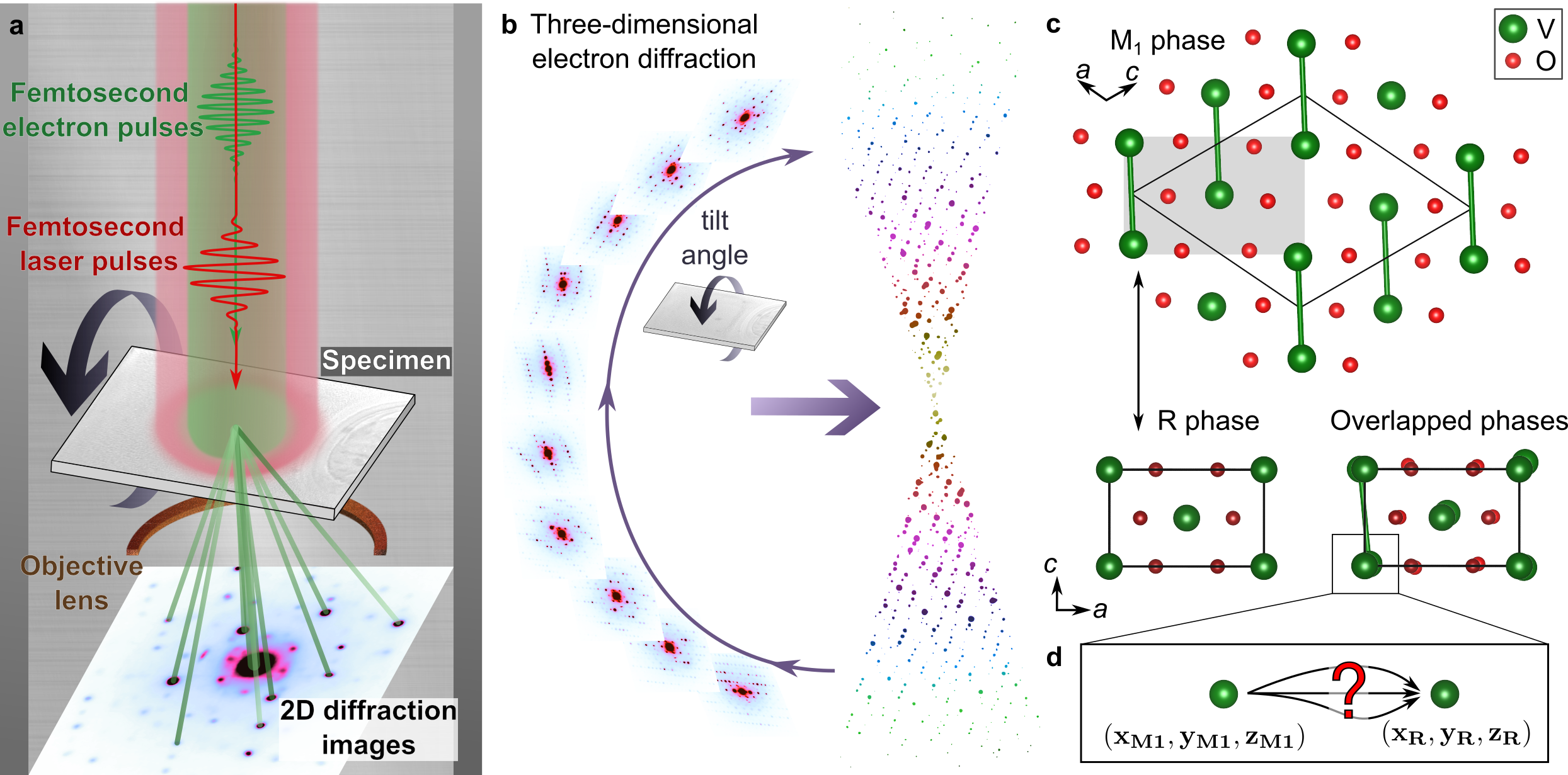}
\caption{\textbf{Ultrafast three-dimensional electron diffraction experiments.} (a) Schematic of the experimental setup. Femtosecond laser pulses induce a structural phase transition in vanadium dioxide, which is probed in reciprocal space using  femtosecond electron pulses with increasing delay. (b) Time-resolved diffraction patterns from multiple tilt angles are combined to construct time-dependent 3D diffraction patterns. (c) $\mathrm{VO_2}$ undergoes a phase transition from the monoclinic (M1) phase to the rutile (R) phase. (d) The precise atomic pathway of the V and O atoms during the monoclinic-to-rutile transition has been debated in the literature for several years~\cite{li_direct_2022,grandi_unraveling_2020}.} \label{fig1}
\end{figure}

\begin{figure}[htbp]%
\centering
\includegraphics[width=0.8\textwidth]{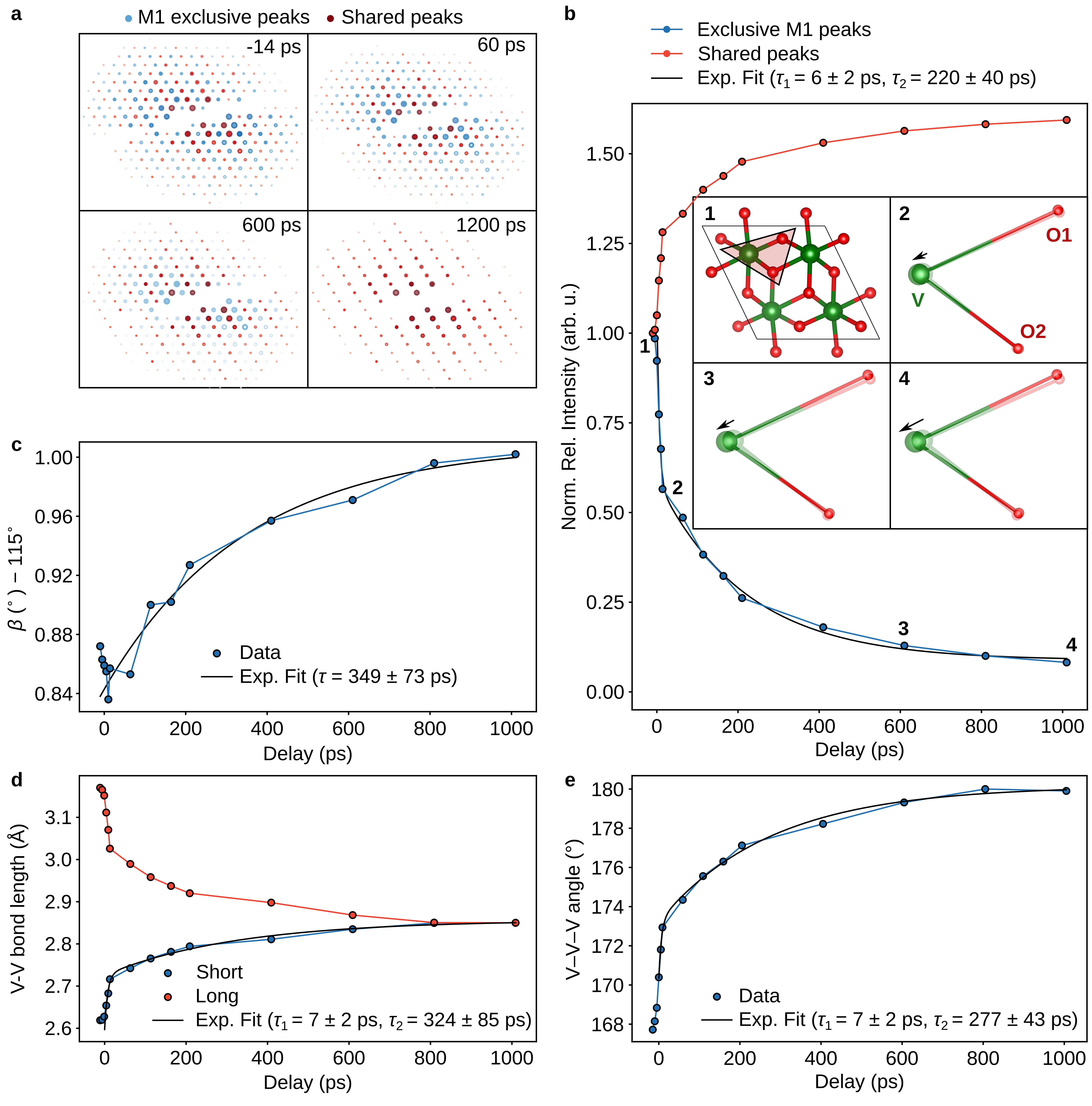}
\caption{\textbf{Structural changes during the light-induced phase transition.} (a) Projection of the three-dimensional electron diffraction pattern viewed along the $b^*$ direction. The Bragg spots which are only present in the monoclinic ground state are highlighted in blue. The diameter of the diffraction discs is proportional to their intensity. (b) The total intensity of the M1 exclusive spots and the shared spots is tracked as a function of the pump-probe delay. A double-exponential fit of the M1 exclusive spots leads to two characteristic time scales. The inset shows the static unit cell (1) for reference along the $b^*$ direction. Asymmetric unit cells (2-4) are presented as a function of the pump-probe delay, with arrows highlighting structural changes over time. The static asymmetric unit cell is also displayed with reduced opacity for comparison. (c) Direct measurements of  the monoclinic beta angle as a function of the pump-probe delay. A characteristic timescale of 380~ps is identified. (d) Vanadium dimerization and (e) zig-zag angle as a function of the pump-probe delay.}\label{fig2}
\end{figure}

\begin{figure}[htbp]%
\centering
\includegraphics[width=0.9\textwidth]{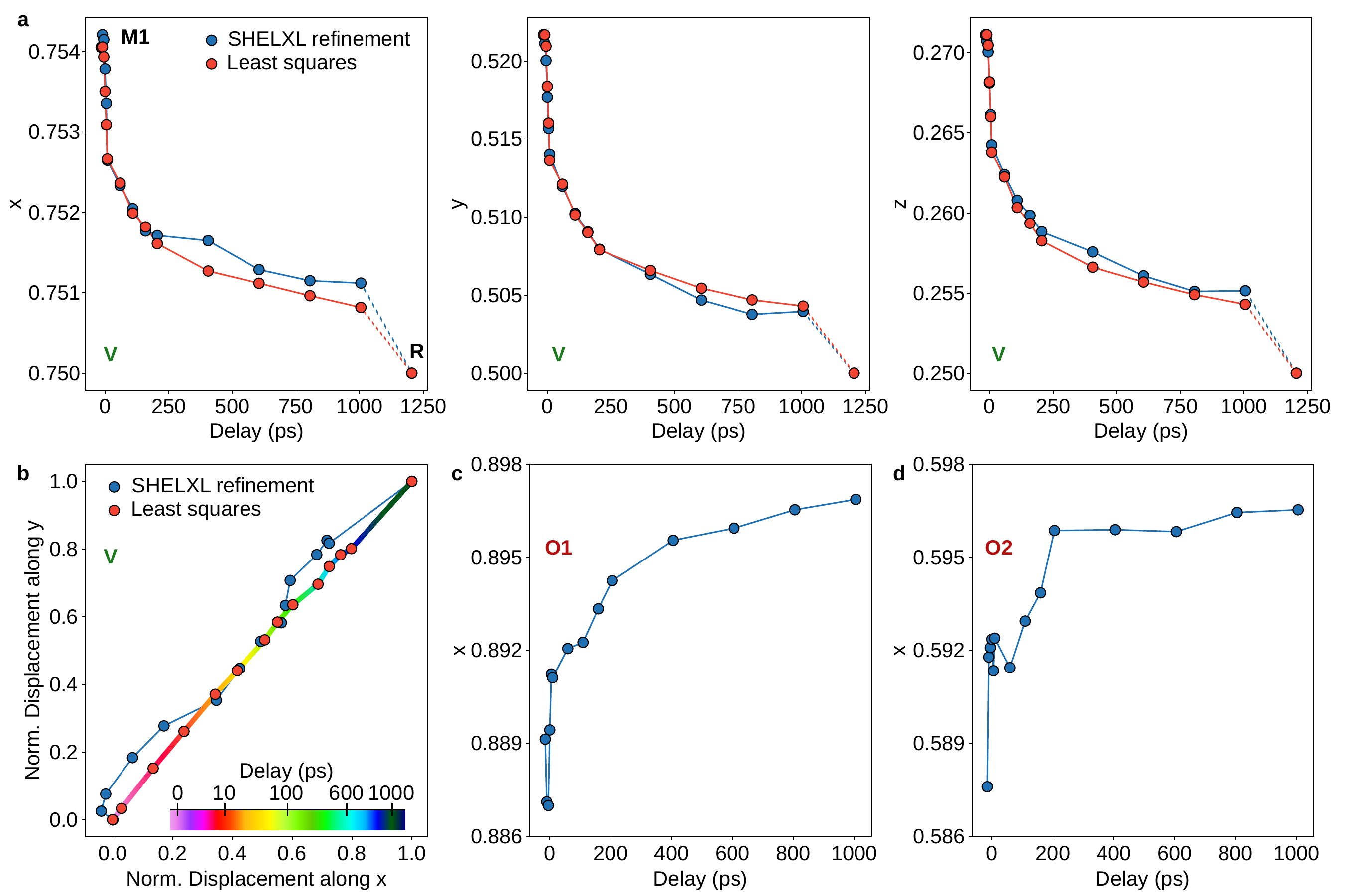}
\caption{\textbf{Atomic fractional coordinates, representing the $\boldsymbol{x, y, z}$ components in real space, calculated from time-dependent three-dimensional diffraction patterns.} (a) The fractional coordinates of the vanadium atom as a function of the pump-probe delay. The results are obtained using the refinement program SHELXL and via least-squares fits of the structure-factor equations. All time points except the final one were indexed in the monoclinic unit cell M1; the final time point was indexed in the rutile unit cell R, but is shown here in the monoclinic representation. Dashed lines indicate the change in indexing. (b) Correlated vanadium motion within the unit cell, showing the relationship between its $x$ and $y$ fractional coordinates. The pathway is direct and linear. (c,d) Fractional coordinates of the two oxygen atoms, O1 and O2, respectively, as a function of the pump-probe delay.}\label{fig3}
\end{figure}

\begin{figure}[htbp]
\centering
\includegraphics[width=0.45\textheight]{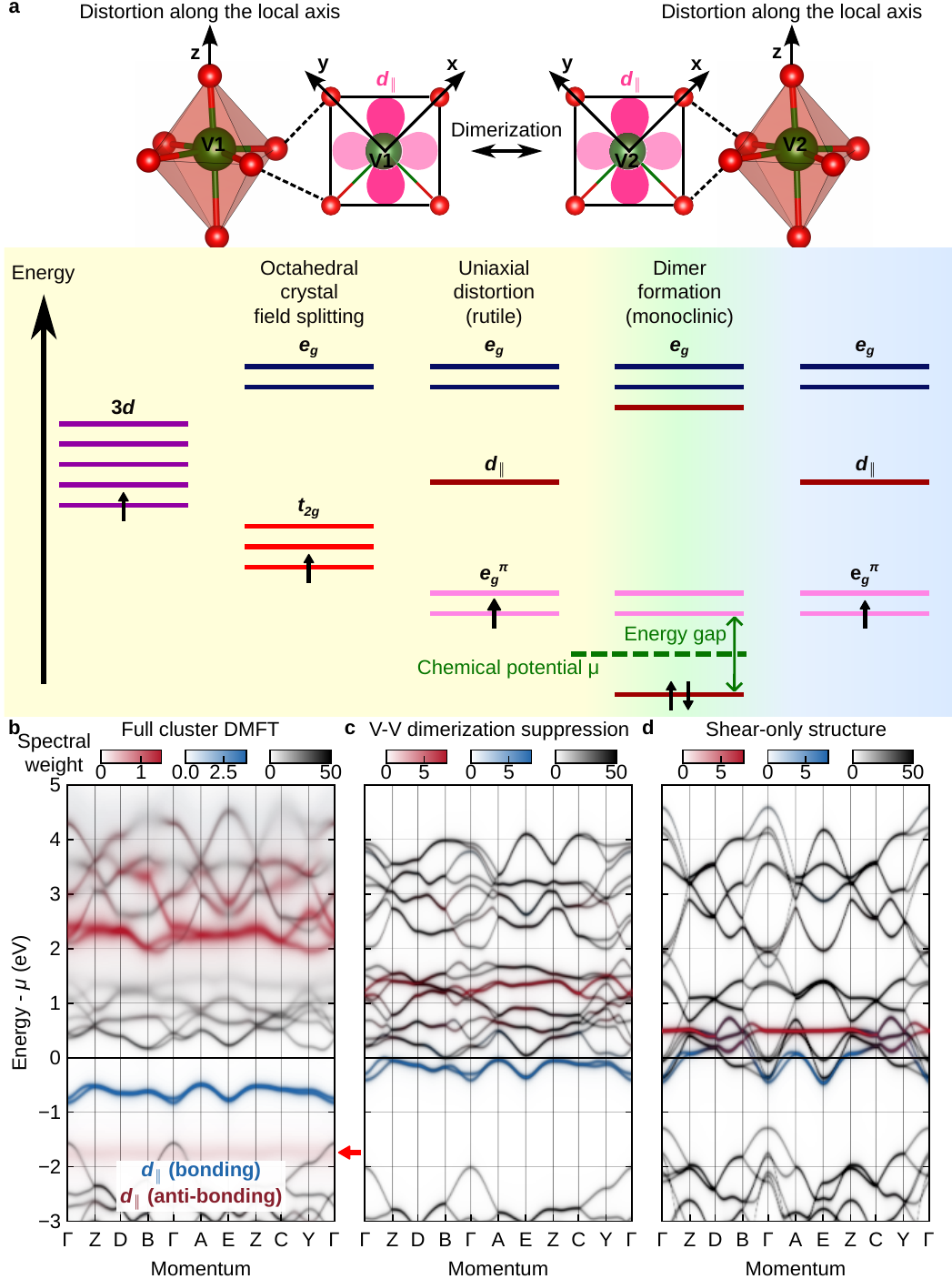}
\caption[Cluster DMFT electronic structure of M1 $\mathbf{VO_2}$ and the
structural origin of the insulating gap]{\textbf{Cluster DMFT electronic structure of M1 $\mathbf{VO_2}$ and the structural
origin of the insulating gap.}
(a) 
Orbital schematic showing the rutile $d_\parallel+e_g^\pi$ manifold and
the monoclinic dimer-split states. V--V dimerization splits the
chain-aligned $d_\parallel$ channel, while $\mathrm{VO_6}$ distortion shifts the
transverse $e_g^\pi$ states.
(b) 
Full cluster DMFT spectral function for refined equilibrium M1 $\mathrm{VO_2}$.
The calculation gives a $0.67$~eV gap and a many-body satellite (red arrow) of the antibonding states $d_\parallel$ (maroon) $1.7$~eV
below $\mu$. The horizontal
dashed line marks $\mu$, which plays the role of the Fermi level at finite temperature.%

 (c) Spectral function for the refined M1 structure, but with $U=J=V_{ex}=0$. Dimerization still produces a sizable one-electron
$d_\parallel$ splitting, but the gap collapses.
(d) Spectral function in the $U=J=V_{ex}=0$ limit for the unit-cell shear-only
structure, with rutile fractional coordinates in the monoclinic cell. The
metallic spectrum shows that monoclinic shear alone does not open a gap.
}
\label{fig:theory_bs}
\end{figure}

\fi

\clearpage 

%
\bibliography{science_template} 
\bibliographystyle{sciencemag}

\end{document}